\documentclass[pra,showpacs,uplatex]{revtex4}

\usepackage{amsmath}
\usepackage{bm}
\usepackage[dvipdfmx]{graphicx}
\usepackage[dvipdfmx]{color}
\usepackage[dvipdfmx]{epsfig}
\usepackage{float}
\usepackage{ulem}
\usepackage{color}

\begin{document}

\title{Superconducting fluctuation current in a space with the Kerr metric}

\author{Hiroshi Kuratsuji}
\affiliation{Department of Physics, Ritsumeikan University-BKC,  Kusatsu,  525--8577, Japan}

\author{Satoshi Tsuchida}
\address{Department of Physics, Osaka City University, 3--3--138 Sugimoto, Sumiyoshi--ku, Osaka City, Osaka, 558--8585, Japan
}

\date{\today}


%
\begin{abstract}
We study a fluctuation current through a superconducting ring placed in a space
with the Kerr metric.  This attempt aims at exploring  conceptual problem;
namely, we are concerned with  the question about the condensed state in an extreme space-time.
The superconducting ring is supposed to be placed in the space surrounding the Kerr black hole (KBH) such that
the rotating axis of the KBH penetrates the center of the ring.
The formulation is based on the Landau-Ginzburg free energy functional of linear form.
The resultant fluctuation current shows several peculiar features that
reflect characteristic aspects of the KBH;
that is, the effect of the Kerr metric gives rise to
the shift of superconducting transition temperature
as well as the scale change of absolute temperature.\\

NOTE: This article has appeared  in Euro Physics Letters, {\bf 126} (2019)10004:

doi: 10.1209/0295-5075/126/10004.
\end{abstract}

\maketitle

{\it 1.Introduction}:
To study the thermodynamical state in the presence of gravitation has long been a fundamental problem  in statistical physics~\cite{LLSP}.
It is also expected that there is a condensed matter counterpart of this problem.
As an early attempt along this thought, we mention the work done
by DeWitt to
evaluate the current caused by the gravitational (frame) drag by utilizing the characteristics of superconductors~\cite{Dewitt}.
This attempt is based on a close resemblance between electromagnetic field
and weak gravity~\cite{LL1,Moller}.
Specifically the gravitational counterpart of a magnetic field should be mentioned.
This is known as the Lense--Thirring field~\cite{LL1,Moller,Zee,Ramos,Mash,Ciufolini},
alias gravitational drag,
which is generated  near a rotating body.
Such a work belongs to the laboratory experiment of gravity~\cite{Caves},
which is in contrast to the detection of the quantum effects
caused by the Earth gravity using, e.g., a neutron interferometer~\cite{Green,Collera}.
The use of a superconductor has an advantage;
huge enhancement of the current would be expected
by using the coherence nature characterizing the superconductivity.
The study of gravitation using the superconductivity has been explored as
a specific category~\cite{Chiao,Tajmar,Anandan}.

In this letter, we address a problem following the same spirit as Ref.~\cite{Dewitt}.
However we consider a full general relativistic effect for the frame dragging
of space-time, which is caused by a rotating black hole known as the Kerr black hole (KBH).
Namely, we suppose that at the  instance when the black hole (BH) is created, the metric suddenly deformed, hence
the current should be induced by the deformation of space-time metric
that is  independent of whether the BH rotates or not.
Once the rotation becomes stationary, the KBH gives rise to a ``gravitational drag" and hence
the stationary current occurs in the ring.

Our object is to evaluate the superconducting current associated with
the deformation of the space-time metric.
Apart from the DeWitt attempt, we are concerned with an estimation of the
{\it fluctuation current} occurring in a superconducting ring such that it is arranged
in a critical condition~\cite{Schmid,Imry,Langer}, by which
one can examine how the superconducting fluctuation should be modified
under an extreme condition of the presence of the KBH.
In this sense, our attempt would be of interest
from a conceptual point of view.

{\it{2. Preliminary}}:
The Kerr metric in the Boyer--Lindquist coordinates is given as~\cite{Boyer}:
\begin{eqnarray}
  ds^{2} = - \left( 1 - \frac{ r_{s} r }{ \Sigma } \right) c^{2} dt^{2}
           - 2 \frac{ r_{s} r a \sin^{2} \theta }{ \Sigma } cdt d{\phi}
           + \frac{ \Sigma }{ \Delta } dr^{2}  \nonumber \\
           + \Sigma d{\theta}^{2}
           + \left( r^{2} + a^{2} + \frac{ r_{s} r a^{2} \sin^{2} \theta }{ \Sigma } \right) \sin^{2} {\theta} d{\phi}^{2}
\end{eqnarray}
where
\begin{eqnarray}
  \Sigma = r^{2} + a^{2} \cos^{2} \theta \ , \ \
  \Delta = r^{2} - r_{s} r + a^{2}   \ , \nonumber \\
  a = \frac{ J }{ M c } \ , \ \
  r_{s} = \frac{ 2 G M }{ c^{2} } \ , \ \
  r_{+} = \frac{ 1 }{ 2 } \left( r_{s} + \sqrt{ r_{s}^{2} - 4 a^{2} } \right)
\end{eqnarray}
Here, $ a $ is the Kerr parameter, $ J $ angular momentum of KBH,
$ M $ mass of the KBH,
$ r_{s} $ Schwarzschild radius,
$ r_{+} $ event horizon,
$ G $ Newtonian constant of gravitation,
and $ c $ is the speed of light.

The starting point is based on an extension of
the time dependent Landau-Ginzburg (LG) theory
to the general relativistic case~\cite{Dinariev,Abrahams}.
In order to realize this
we use the action for the complex scalar field $ {\varphi} $~\cite{LL1,Weinberg},
that is given as
\begin{eqnarray}
  \label{eq:csaction}
  S = \frac{ \hbar }{c} \int \sqrt{ - g } \left[ g^{ \mu \nu } \left( {\partial}_{\mu} {\varphi}^{*} \right) \left( {\partial}_{\nu} {\varphi} \right)
                                  - V \left( {\varphi}, {\varphi}^{*} \right) \right] d^{4}x
\end{eqnarray}
where $ g $ is the determinant of
the metric tensor $ g_{ \mu \nu} $.
$ V ( {\varphi}, {\varphi}^{*} ) $ is the potential,
which is assumed to have the form:
\begin{equation}
 V = - \frac{ c^{2} }{ {\hbar}^{2} } {\mu}^{2} {\varphi}^{*} {\varphi}
\equiv  - \left[ \frac{ c^{2} }{ {\hbar}^{2} } (2m_e)^{2} + F(T)  \right] \varphi^{*} \varphi
\label{mu}
\end{equation}
where $ m_{e} $ is the electron mass,
the factor 2 implies the Cooper pair.
$ F(T) $, which represents
a thermodynamic quantity,
describes the feature near
the superconducting transition temperature
$ T = T_{c} $,
and the explicit form will be given later.

\begin{figure}[b]
  \begin{center}
      \includegraphics[width=40mm]{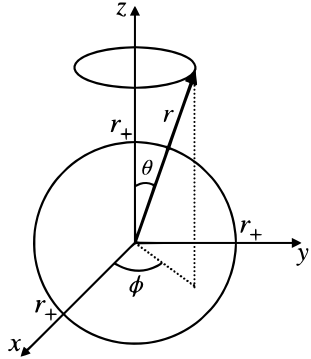}
    \caption{(Color Online).
      The setting for estimating the current caused by
      gravitational drag induced by a KBH.
      The coordinate $ r $ is the observational point,
      $ {\theta} $ is the zenith angle,
      and $ r_{+} $ is the outer horizon of the KBH.}
   \label{fig:drag1}
  \end{center}
\end{figure}

We restrict the expression of the action function for the complex scalar field
in a special geometry for the two dimensional space-time;
we consider a superconducting circular ring placed above the rotating black hole (KBH)
such that the ring is coaxial to the KBH, and the radius is prescribed to be smaller than
the radius of the KBH as shown in Fig.~\ref{fig:drag1}.
We adopt an ideal limit; the thickness of the ring is neglected,
that is, the radius is extremely larger than the cross section of the ring.
In this geometry, we introduce the order parameter, $\psi$,
which is assumed to depend on
angular variable $ {\phi} $ and time $ t $,
namely $ \psi = \psi( \phi, t ) $,
then the action function can be written as:
\begin{eqnarray}
  \label{eq:opaction}
  S = \frac{\hbar}{c} \int \left[ A \frac{ {\partial} {\psi}^{*} }{ {\partial} t } \frac{ {\partial} {\psi} }{ {\partial} t }
        + B \left( \frac{ {\partial} {\psi}^{*} }{ {\partial} t } \frac{ {\partial} {\psi} }{ {\partial} \phi }
                      + \frac{ {\partial} {\psi} }{ {\partial} t } \frac{ {\partial} {\psi}^{*} }{ {\partial} \phi } \right) \right. \nonumber \\
      \left. + ~C \frac{ {\partial} {\psi}^{*} }{ {\partial} \phi } \frac{ {\partial} {\psi} }{ {\partial} \phi }
        + \sqrt{ - g } \frac{ c^{2} }{ {\hbar}^{2} } {\mu}^{2} {\psi}^{*} {\psi} \right] d {\phi} dt
\end{eqnarray}
Here, $ \sqrt{ -g } = c \Sigma \sin \theta $,
and $ A $, $ B $, $ C $ are given as follows:
\begin{eqnarray}
  \label{eq:ABC}
  A &=& \sqrt{ -g } g^{tt} = - \frac{ {\Sigma} \sin \theta }{ c \Delta }
                              \left( r^{2} + a^{2} + \frac{ r_{s} r a^{2} {\sin}^{2} \theta }{ \Sigma } \right)  \nonumber \\
  B &=& \sqrt{ -g } g^{t {\phi}} = \sqrt{ -g } g^{{\phi}t}
     =  - \frac{ r_{s} r a \sin \theta }{ \Delta }  \\
  C &=& \sqrt{ -g } g^{ {\phi} {\phi} }
     =  \frac{ c \Sigma }{ \Delta \sin \theta } \left( 1 - \frac{ r_{s} r }{ \Sigma } \right)  \nonumber
\end{eqnarray}
Now, we introduce the Lagrandian density $ {\mathcal{L}} $ as
$ S = \int {\mathcal{L}} d{\phi} dt $, that is,
\begin{eqnarray}
  \label{eq:lagden}
  {\mathcal{L}} = \frac{\hbar}{c} \left[ A \frac{ {\partial} {\psi}^{*} }{ {\partial} t } \frac{ {\partial} {\psi} }{ {\partial} t }
        + B \left( \frac{ {\partial} {\psi}^{*} }{ {\partial} t } \frac{ {\partial} {\psi} }{ {\partial} \phi }
                      + \frac{ {\partial} {\psi} }{ {\partial} t } \frac{ {\partial} {\psi}^{*} }{ {\partial} \phi } \right) \right. \nonumber \\
        \left. + ~C \frac{ {\partial} {\psi}^{*} }{ {\partial} \phi } \frac{ {\partial} {\psi} }{ {\partial} \phi }
        + \sqrt{ -g } \frac{ c^{2} }{ {\hbar}^{2} } {\mu}^{2} {\psi}^{*} {\psi} \right]
\end{eqnarray}
The second term characterized by the coefficient $ B $
describes the {\it dragging effect},
which is proportional to the Kerr parameter $ a $.
As is well known,
this gives rise to the Lense--Thirring field in the linear approximation limit.

In accordance with the general procedure in quantum mechanics,
we can separate the order parameter into time and space-dependent parts.
Hence, to obtain the stationary LG form,
it is natural to adopt the following time-dependence for the order parameters:
\begin{eqnarray}
  \label{eq:op}
  \psi ( \phi, t ) = e^{ i \omega t } \tilde{ \psi } ( \phi )
  \ \ , \ \
  \psi^{*} ( \phi, t ) = e^{ - i \omega t } \tilde{ \psi }^{*} ( \phi ),
\end{eqnarray}
where $ \omega $ has the dimension of frequency,
which should satisfy a condition that will be determined later.
Then, the Lagrangian density $ {\mathcal{L}} $ becomes
\begin{eqnarray}
  {\mathcal{L}} = \frac{\hbar}{c} \left[ C \frac{ {\partial} \tilde{{\psi}}^{*} }{ {\partial} \phi } \frac{ {\partial} \tilde{{\psi}} }{ {\partial} \phi }
                  + i \omega B \left( \tilde{\psi} \frac{ {\partial} \tilde{{\psi}}^{*} }{ {\partial} \phi } - \tilde{\psi}^{*} \frac{ {\partial} \tilde{{\psi}} }{ {\partial} \phi } \right) \right. \nonumber \\
                  \left. + \left( {\omega}^{2} A + \sqrt{ -g } \frac{ c^{2} }{ {\hbar}^{2} } {\mu}^{2} \right) \tilde{\psi}^{*} \tilde{\psi} \right]
\end{eqnarray}
Thus the LG free energy $ F_{g} $ is given by
\begin{eqnarray}
  F_{g} = \frac{ 1 }{ \tau } \int {\mathcal{L}} d {\phi} dt = \frac{ 1 }{ \tau } S
 \end{eqnarray}
with $ \tau = \frac{ 2 \pi }{ \omega } $,
and this turns out to be
\begin{eqnarray*}
  \label{eq:fg}
  F_{g} &=& \frac{\hbar}{c}
          \int \tilde{\psi}^{*} \left[ C \left( -i \frac{ \partial }{ {\partial} {\phi} } + \omega \frac{ B }{ C } \right)^{2} \right. \nonumber \\
          && \ \   \left. + \left\{ {\omega}^{2} \left( A - \frac{ B^{2} }{ C } \right) + \sqrt{ -g } \frac{ c^{2} }{ {\hbar}^{2} } {\mu}^{2} \right\}
                \right] \tilde{\psi} d {\phi}
\end{eqnarray*}

{\it{3. Fluctuating current}}:
Now, we calculate the partition function using the LG free energy,
which is given by the functional integral:
\begin{equation}
  \label{eq:partfunc}
  Z = \int \exp \left[ -\beta F_{g} \right] \mathcal{D} \left[ \tilde{\psi}, \tilde{\psi}^{*} \right]
\end{equation}
The LG free energy can be expressed in terms of the linear LG theory~\cite{Schmid,LP}:
$ F_{g} = \int \psi^{*}\Lambda \psi d \phi $,
where $ \Lambda $ is given by
\begin{eqnarray}
  \label{eq:lambdahami}
  \Lambda &=&
  \frac{ \hbar }{ c } C \left( -i \frac{ d }{ d \phi } + \omega \frac{B}{C} \right)^{2}
  + \vert \alpha \vert,  \nonumber \\
  \alpha &\equiv& \frac{ \hbar }{ c } \left[ {\omega}^{2} \left( A - \frac{ B^{2} }{ C } \right) + \sqrt{ -g } \frac{ c^{2} }{ {\hbar}^{2} } {\mu}^{2} \right]
\end{eqnarray}
Let us expand the order parameter as $ \tilde{\psi} ( \phi ) = \sum_{n} a_{n} \tilde{\psi}_{n} (\phi) $.
Here $ \tilde{\psi}_{n} $ are the eigenfunctions for the eigenvalue equations,
$ \Lambda \tilde{\psi}_{n} = \lambda_{n} \tilde{\psi}_{n} $,
with $ n $ taking an  integer value running from $ - \infty $ to $ + \infty $.
Taking account of the periodic boundary condition $ \tilde{\psi} (\phi) = \tilde{\psi} (\phi + 2n \pi) $,
one can write $ \tilde{\psi} (\phi) \propto e^{i n \phi } $,
the eigenvalues $ {\lambda}_{n} $ are calculated to be
\begin{eqnarray}
  \lambda_{n} = \frac{ \hbar }{ c } C \left( n + \omega \frac{B}{C} \right)^{2} + \vert \alpha \vert
  \label{eq:smalllambda}
\end{eqnarray}
and the corresponding eigenfunctions become $ \tilde{\psi}_{n} (\phi) = \frac{ 1 }{ \sqrt{2 \pi} } \exp \left[ i n \phi \right] $.
The partition function is thus calculated by using the functional integral in Eq.~(\ref{eq:partfunc}):
\begin{eqnarray}
  Z &=& \int \exp \left[ - \beta \sum_{n} \lambda_{n} a_{n}^{*} a_{n} \right] \prod_{n} da_{n}^{*} da_{n} \nonumber \\
  &=& \prod_{ n = - \infty }^{ + \infty } \frac{ 2 \pi }{ \beta \lambda_{n} }
  \label{partition}
\end{eqnarray}
from which, up to some additional constant,
one obtains the free energy $ F = - k_{B} T \log Z $;
\begin{eqnarray}
  F &=& k_{B} T \sum_{n} \log \left[ \left( n + \omega \frac{B}{C} \right)^{2} + \vert \gamma \vert \right] \nonumber \\
  &=& k_{B} T \log \left[ \cosh ( 2 \pi \sqrt{\gamma} ) - \cos \left( 2 \pi \omega \frac{B}{C} \right) \right]
\end{eqnarray}
Here we adopt the notation $  \gamma =  \frac{ c }{ \hbar } \frac{ \alpha }{ C } $.
The quantity proportional to $ \frac{B}{C} $ just plays the same role as  the
vector potential penetrating the ring,
hence following the procedure used in the mesoscopic physics~\cite{Imry},
we rewrite it in  the form
\begin{equation}
 \omega \frac{B}{C} = \frac{\Phi}{\phi_0}
\end{equation}
Here $ \phi_0 $ stands for the flux quantum for the Cooper pair;
$ \phi_0 = \frac{hc}{2e} $.
Using this, the current can be concisely evaluated by the formula:
$ I = c\frac{\partial F}{\partial \Phi} $, which leads to
\begin{equation}
  I = \frac{ 2 \pi c k_{B} T }{ {\phi}_{0} }
      \frac{ \sin \left( \frac {  2 \pi \Phi }{ \phi_{0} } \right) }{ \cosh ( 2 \pi \sqrt{\gamma} ) - \cos \left( 2 \pi \frac{\Phi}{\phi_0} \right) }
\end{equation}
Here we have used the formula given in Ref.~\cite{math}.
Thus, the current is a periodic function with the period $ \phi_0 $.
Noting this feature, we can estimate the average current
by averaging over the half period $ \phi_0/2 $.
Then we obtain the average current:
\begin{eqnarray}
  \bar{I} &=& \frac{ 1 }{ \phi_{0} / 2 } \int_{0}^{ \phi_{0} / 2 } I d {\Phi}  \nonumber \\
          &=& \frac{ 4 c k_{B} T }{ \phi_{0} }
            \log \left[ \cosh \left( \pi \sqrt{\gamma} \right) - \sinh \left( \pi \sqrt{\gamma} \right) \right]
\end{eqnarray}
This shows a critical behavior in the region where $ \gamma \sim 0 $:
\begin{equation}
  \bar I \sim \log \sqrt{\gamma}
\end{equation}
namely,
the current shows the $ \log $ divergences.
In what follows, we shall examine the physical contents of this peculiar
behavior.

{\it{4. Discussions}}:
Here, the thermodynamical quantity $ F(T) $ in Equation (\ref{mu})
can be chosen as $ F(T) = K(T-T_c) $ (see Refs.~\cite{Schmid, LP}),
where $ K $ is the intrinsic quantity of the superconducting material,
and $ T_{c} $ is the superconducting transition temperature.
Then, $ \alpha $ is written as
\begin{eqnarray}
  \label{eq:alpha}
  \alpha &=& \frac{ \hbar }{ c } \left[ {\omega}^{2} \left( A - \frac{ B^{2} }{ C } \right) +
  \sqrt{ -g } \left\{ \frac{ c^{2} }{ {\hbar}^{2} } (2m_e)^{2} + K(T-T_c)  \right\} \right]  \nonumber \\
  &\equiv& \frac{ \hbar }{ c } \sqrt{ -g } K ( T - T_{c}')
\end{eqnarray}
At this point, we need to fix the frequency $ \omega $,
which is left as an undetermined parameter.
This procedure is given as follows:
In the ordinary flat space (Minkowski space),
the factor $ \alpha $ should turn out to be
$ \alpha = \frac{ \hbar }{ c } K ( T - T_{c }) $,
hence we can obtain $ \omega = \frac{ 2 m_{e} c^{2} }{ \hbar } $.

\begin{figure}[b]
 \hspace{-15mm}
  \includegraphics[width=90mm]{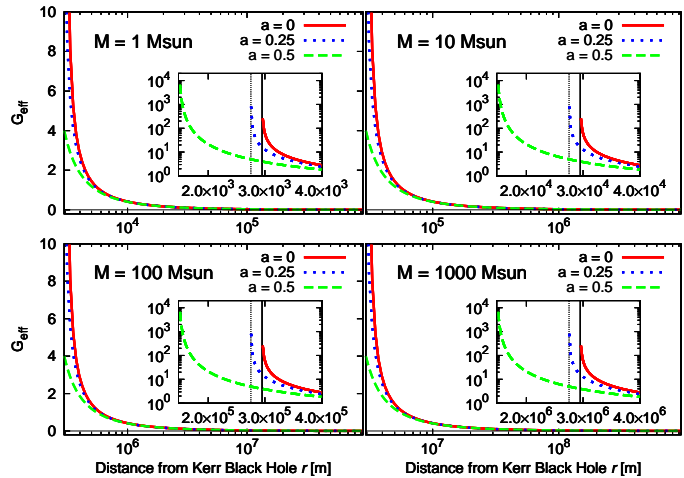}
  \vspace{-10mm}
  \begin{center}
    \caption{(Color Online).
    The value of $ G_{\rm{eff}} $ as a function of
    the distance from the Kerr black hole $ r $.
    We choose four cases of the KBH mass;
    $ M = $ 1 (left top), 10 (right top), 100 (left bottom), 1000 (right bottom)~$ M_{\odot} $.
    For each panel, the solid line, dotted line, and dashed line show
    the case for $ a = $~0, 0.25, and 0.5 multiplied by $ r_{s} $, respectively.
    For each small panel, the solid and dotted vertical lines indicate
    the event horizons of KBH for $ a = 0 $ and $ a = 0.25~r_{s} $, respectively.
 }
    \label{fig:grav_eff}
  \end{center}
\end{figure}

From the above derivation,
we arrive at the main two consequences.
First, the concrete expression for
$ \gamma $, which determines the average current,
can be obtained in the form:
\begin{eqnarray}
  \label{eq:mod_gamma}
  \gamma = \frac{c}{\hbar} \frac{\alpha}{C} = \sqrt{ -g } K \frac{T-T_c'}{C}
\end{eqnarray}
This implies the scale change for the absolute temperature,
which is induced by the metric tensor proportional to $ g^{ {\phi}{\phi} } $ explicitly.
More precisely, the modified transition temperature $ T_{c}' $,
as is seen below,
is also scaled by the metric tensor $ g^{ t \phi } $ and $ g^{ t t } $.
This consequence is consistent with the well-known result
(see, e.g, section 27 in Ref.~\cite{LLSP}).
At the transition point, $ T = T_{c}' $,
$ \gamma $ becomes zero as well as $ \alpha $,
which causes the divergence of the current.
The second concerns  the effective transition temperature $ T_{c}' $,
which is writtten as
\begin{eqnarray}
  \label{eq:mod_tc}
  T_c' &=&  T_{c} + \frac{ 1 }{ K } \left( \frac{ 2 m_{e} c }{ \hbar } \right)^{2} \left[ \frac{ c^{2} }{ \sqrt{ -g } } \left( \frac{ B^{2} }{ C } - A \right) - 1 \right]  \nonumber \\
       &=& T_{c} + {\Delta} T_{c}
\end{eqnarray}
This represents the shift of the transition temperature.
%
Here, the shift, $ \Delta T_{c} $, is always non negative.
%
The factor
$ \frac{ 1 }{ K } \left( \frac{ 2 m_{e} c }{ \hbar } \right)^{2} $
is a quantity intrinsic to matter,
whereas the factor [...],
which is denoted as $ G_{\rm{eff}} $,
indicates the gravitational effect caused by the Kerr metric.
The latter effect is shown in Fig.~\ref{fig:grav_eff}.
For each panel, the solid line, dotted line, and dashed line show
the case for $ a = $~0, 0.25, and 0.5
multiplied by $ r_{s} $, respectively.
For each small panel, the solid and dotted vertical lines indicate
the event horizons of KBH for $ a = 0 $ and $ a = 0.25~r_{s} $, respectively.
This figure shows that near the event horizon of KBH,
that is $ r \sim r_{+} $,
the shift of temperature $ \Delta T_{c} $ is extremely enhanced.
Furthermore, Fig.~\ref{fig:grav_eff} indicates that
$ \Delta T_{c} $ has a tendency to increase
with the  increase of  the Kerr parameter $ a $.
In addition, we can see from Fig.~\ref{fig:grav_eff} that
$ G_{\rm{eff}} $ tends to zero
when the distance $ r $ becomes sufficiently large.
Thus, the effect caused by this factor does not appear in the Minkowski metric,
whereas the effect will be extracted dramatically in strong gravitational field.
We note that, the effect of varying $ \theta $ can be negligible
for  a small value of $ \theta $,
so we do not take account of a difference of $ \theta $ in this figure.
We choose four cases of the KBH mass;
$ M = $ 1, 10, 100, 1000~$ M_{\odot} $.
As is intuitively expected,
the qualitative behavior of $ G_{\rm{eff}} $ does not change for any KBH mass.
On the other hand, the distance to which the gravitational effect extends,
will become large with the increase of  the KBH mass.
%

Here, we give the physical interpretation concerning the increase of the transition temperature.
This feature implies that the condensation of a Cooper pair in the curved space
looks like to be made more easily  than in the Minkowski space.
This consequence seems to be counterintuitive
from the point of ordinary condensation in the Minkowski space.
However, one could rather interpret this situation as follows:
The condensed state of a  Cooper pair would survive
even in the high temperature region
due to the effect of the Kerr metric.
In this connection,
we mention that,
in the one dimensional superconductor,
with which we are concerned here,
it is known that the transition temperature becomes broad
due to the effect of fluctuation~\cite{Imry}.
We expect that some implications may be given from this phenomenon,
although the present gravity effect may not have direct relation with it.

{\it {5. Summary}}:
We have explored the gravitational effect for the condensed matter state by
exemplifying the general relativistic LG theory for the superconducting ring in the Kerr metric.
The resultant fluctuation current concisely shows up a drastic change of
the transition temperature as well as the scale change of the temperature itself caused by the Kerr metric.
There may remain a related problem:
the  occurrence of the transient process from the static black hole to the final stationary rotating black hole, which
may be regarded as a sort of the electromagnetic induction in the electromagnetic theory.


\newpage

\begin{thebibliography}{99}

  \bibitem{LLSP}
  L. D. Landau and E. M. Lifschitz,
  {\it{Statistical Physics,
  Course of Theoretical Physics}},
  Vol. 5 (Butterworth-Heinemann, Oxford, 1984).

  \bibitem{Dewitt}
  B. S. DeWitt,
  Phys. Rev. Lett. {\bf{16}}, 1092 (1966).

  \bibitem{LL1}
  L. D. Landau and E. M. Lifshitz,
 { \it{ Classical Theory of Fields,
  Course of Theoretical Physics}},  Vol. 2, fourth edition,
  (Butterworth--Heinemann, 1980).

  \bibitem{Moller}
  C. M$\phi$ller,
  {\it{The Theory of Relativity}},
  (Oxford University Press, 1952).

  \bibitem{Zee}
  A. Zee,
  Phys. Rev. Lett. {\bf{55}}, 2379 (1985).

  \bibitem{Ramos}
  J. Ramos and B. Mashhoon,
  Phys. Rev. D {\bf{73}}, 084003 (2006);

  \bibitem{Mash}
  B. Mashhoon, in
  {\it{Reference Frames and Gravitomagnetism}},
  edited by J. F. Pascual--S{\'a}nchez, L. Floria,
  A. San Miguel, F. Vicente
  (World Scientific, Singapore, 2001).

  \bibitem{Ciufolini}
  I. Ciufolini,
  Nature {\bf{449}}, 41 (2007).

  \bibitem{Caves}
  V. B. Braginsky, C. M. Caves and K. S. Thorne,
  Phys. Rev. D {\bf{15}}, 2047 (1977).

  \bibitem{Green}
  D. Greenberger,
  Ann. Phys. {\bf{47}}, 116 (1968);
  D. Greenberger and A. W. Overhauser,
  Rev. Mod. Phys. {\bf{51}}, 43 (1979).

  \bibitem{Collera}
  R. Colella, A. W. Overhauser, and S. A. Werner,
  Phys. Rev. Lett. {\bf{34}}, 1472 (1975).

  \bibitem{Chiao}
  S. J. Minter, K. W. McNelly, and R. Y. Chiao,
  Physica E, {\bf{42}}, 234 (2010).

  \bibitem{Tajmar}
  M. Tajmar, C. J. de Matos,
  Physica C {\bf{420}}, 56 (2005).

  \bibitem{Anandan}
  J. Anandan,
  Phys. Rev. Lett. {\bf{47}}, 463 (1981).

  \bibitem{Schmid}
  A. Schmid,
  Phys. Rev. {\bf{180}}, 527 (1969).

  \bibitem{Imry}
  Y. Imry,
  {\it{Introduction to Mesoscopic physics}}, 2nd ed.
  (Oxford Univercity Press, 2002).

  \bibitem{Langer}
  J. S. Langer, V. Ambegaokar
  Phys. Rev. {\bf{164}}, 498 (1967).

  \bibitem{Boyer}
  R. H. Boyer, R. W. Lindquist
  Journal of Math. Phys. {\textbf{8}}, 265 (1967).

  \bibitem{Dinariev}
  O. Y. Dinariev and A. B. Mosolov,
  Sov. Phys. J. {\textbf{32}} , 315 (1989).

  \bibitem{Abrahams}
  H. Suhl,
  Phys. Rev. Lett. {\textbf{14}}, 226 (1965);
  E. Abrahams and T. Tsuneto,
  Phys. Rev. {\textbf{152}}, 416 (1966).

  \bibitem{Weinberg}
  S. Weinberg,
  {\it{Cosmology}},
  (Oxford University Press, 2008).

  \bibitem{LP}
  E. M. Lifshitz and L. P. Pitaevski,
  {\it Statistical Physics; part 2},
  Course of Theoretical Physics, vol.~9.
  (Butterworth-Heinemann, 1986).

  \bibitem{math}
  K. Ito,
   {\it Encyclopedic Dictionary of Mathematics}, 4th ed.

  (MIT Press, Cambridge, 1993);
  H. Ono,  Private communication.

\end{thebibliography}
\end{document}